\documentclass[aip,amsmath,amssymb,preprint]{revtex4-1}
\usepackage{upgreek}
\usepackage{graphicx}
\usepackage{dcolumn}
\usepackage{bm}

\begin{document}

\title[]{Screening of the quantum-confined Stark effect in AlN/GaN nanowire superlattices by Germanium doping}

\author{P.~Hille}
 \email{Pascal.Hille@physik.uni-giessen.de}
 \affiliation{I. Physikalisches Institut, Justus-Liebig-Universit\"at Gie\ss en, Heinrich-Buff-Ring 16, 35392 Gie\ss en, Germany}
\author{J.~M\"{u}{\ss}ener}%
 \affiliation{I. Physikalisches Institut, Justus-Liebig-Universit\"at Gie\ss en, Heinrich-Buff-Ring 16, 35392 Gie\ss en, Germany}
\author{P.~Becker}
 \affiliation{I. Physikalisches Institut, Justus-Liebig-Universit\"at Gie\ss en, Heinrich-Buff-Ring 16, 35392 Gie\ss en, Germany}
\author{M.~de la Mata}
 \affiliation{Institut de Ciencia de Materials de Barcelona, ICMAB-CSIC, Campus de la UAB, 08193 Bellaterra, CAT, Spain}
\author{N.~Rosemann}
 \affiliation{Faculty of Physics and Materials Science Center, Philipps Universit\"at Marburg, Renthof 5, 35032 Marburg, Germany}
\author{C.~Mag\'{e}n}
 \affiliation{Laboratorio de Microscopías Avanzadas, Instituto de Nanociencia de Aragon-ARAID, Universidad de Zaragoza, 50018 Zaragoza, Spain}
\author{J.~Arbiol}
 \affiliation{Institut de Ciencia de Materials de Barcelona, ICMAB-CSIC, Campus de la UAB, 08193 Bellaterra, CAT, Spain}
 \affiliation{Institucio Catalana de Recerca i Estudis Avançats (ICREA), 08010 Barcelona, CAT, Spain}
\author{J.~Teubert}
 \affiliation{I. Physikalisches Institut, Justus-Liebig-Universit\"at Gie\ss en, Heinrich-Buff-Ring 16, 35392 Gie\ss en, Germany}
\author{S.~Chatterjee}
 \affiliation{Faculty of Physics and Materials Science Center, Philipps Universit\"at Marburg, Renthof 5, 35032 Marburg, Germany}
\author{J.~Sch\"{o}rmann}
 \affiliation{I. Physikalisches Institut, Justus-Liebig-Universit\"at Gie\ss en, Heinrich-Buff-Ring 16, 35392 Gie\ss en, Germany}
\author{M.~Eickhoff}
 \affiliation{I. Physikalisches Institut, Justus-Liebig-Universit\"at Gie\ss en, Heinrich-Buff-Ring 16, 35392 Gie\ss en, Germany}

\date{\today}

\begin{abstract}
We report on electrostatic screening of polarization-induced internal electric fields in AlN/GaN nanowire heterostructures with Germanium-doped GaN nanodiscs embedded between AlN barriers. The incorporation of Germanium at concentrations above 10$^{20}$\,cm$^{-3}$ shifts the photoluminescence emission energy of GaN nanodiscs to higher energies accompanied by a decrease of the photoluminescence decay time. At the same time, the thickness-dependent shift in emission energy is significantly reduced. In spite of the high donor concentration a degradation of the photoluminescence properties is not observed.
\end{abstract}

\maketitle

Reports on the self-assembled growth of GaN nanowires (NWs) by plasma-assisted molecular beam epitaxy (PAMBE) more than a decade ago \cite{Yoshizawa1997,Calleja2000} kindled the interest in this promising technology platform for the realization of light emitters with improved efficiency \cite{Kikuchi2004,Armitage2010}. More recently, applications such as photocatalytic water splitting \cite{Jung2008,Wang2011,Benton2013} or optochemical sensing have been proposed \cite{Teubert2011,Wallys2012b}. III-N NWs exhibit a much lower density of structural defects compared to planar thin films/structures due to their three-dimensional geometry and the corresponding efficient strain-relaxation \cite{Calleja2001,Cerutti2006}. Hence, these structures are also considered to be a model system for the analysis and relation of basic structural and optical properties \cite{Arbiol2009,Li2012,DeLaMata2013} in addition to their huge application potential. The realization of Al$_\text{x}$Ga$_\text{1-x}$N/GaN NW heterostructures (NWHs) \cite{Ristic2003} triggered the analysis of carrier confinement mechanisms in NWHs with embedded GaN nanodiscs (NDs). Besides the influence of the specific NW geometry via the so-called “strain-confinement” \cite{Ristic2005,Rivera2007}, the large polarization-induced internal electrical fields for Al$_\text{x}$Ga$_\text{1-x}$N barriers with high Al-content exert strong influence on the optical properties of NWHs via the quantum-confined Stark effect (QCSE). As a result, the photoluminescence (PL) emission energy of NDs embedded in Al$_\text{x}$Ga$_\text{1-x}$N barrier material has been shown to be shifted below the GaN band gap for a sufficiently large Al-content and disc thickness \textit{d}$_\text{ND}$ \cite{Renard2009,Furtmayr2011}, as it has been reported before for planar Al$_\text{x}$Ga$_\text{1-x}$N/GaN quantum well (QW) structures \cite{Leroux1998,Grandjean1999}. The possibility of screening the polarization-induced internal electric fields in planar Al$_\text{x}$Ga$_\text{1-x}$N/GaN QWs by doping has been theoretically investigated in Ref.\cite{DiCarlo2000}. Experimentally, screening by photoexcited carriers after excitation with different excitation powers was observed in Al$_\text{x}$Ga$_\text{1-x}$N/GaN QWs \cite{Alderighi2001,Vinattieri2002,Fujita2007} as well as in AlN/GaN quantum dots (QDs) \cite{Kalliakos2004,Widmann1998}. 
Si-doping of the active QD-region has been reported not to affect the internal electric fields in AlN/GaN QD layers \cite{Guillot2006} most likely due to the limited doping efficiency of Si and its deteriorating influence on the crystal quality when incorporated in high concentrations \cite{Forghani2012}. In Ref.\cite{Haratizadeh2004} even a red shift of the PL emission of Al$_\text{0.07}$Ga$_\text{0.93}$N/GaN QWs due to Si-doping of the QWs was observed.
Recently, efficient n-type doping of GaN layers \cite{Hageman2004,Fritze2012,Kirste2013} and GaN NWs \cite{Schoermann2013} using Germanium as a donor was demonstrated and Ge-concentrations above 10$^{20}$\,cm$^{-3}$ could be achieved without structural degradation of the material, opening the possibility for screening of the QCSE by free carriers.\\ 
In this work, we report on the electrostatic screening of internal electric fields in AlN/GaN ND superlattices (NDSLs) by free carriers due to Ge-doping of the GaN NDs. An increasingly suppressed QCSE with increasing Ge-concentration is demonstrated by a combination of continuous-wave (cw-) and time-resolved (TR-)PL analysis. 

Growth of N-polar III-N NWHs was carried out by PAMBE using nitrogen-rich growth conditions on highly conductive n-Si(111) substrates \cite{Furtmayr2011,delaMata2012}. After growth of a 600\,nm nominally undoped GaN NW base a NDSL structure containing 40 GaN NDs (ND thickness \textit{d}$_\text{ND}$\,=\,1.6\,--\,7.7\,nm) embedded in AlN barriers (nominally 4\,nm) was grown and covered with a 20\,nm GaN cap. The Ga beam equivalent pressure (BEP$_\text{Ga}$) was kept constant at 3\,x\,10$^{-7}$\,mbar leading to a growth rate of 6.5\,nm/min. The AlN barriers were grown with a BEP$_\text{Al}$ of 8\,x\,10$^{-8}$\,mbar resulting in a growth rate of 2.8\,nm/min. Germanium was used as n-type dopant of the GaN NDs and the BEP$_\text{Ge}$ was varied between 5\,x\,10$^{-10}$ and 3\,x\,10$^{-9}$\,mbar. Linear extrapolation of the data reported for GaN NWs in Ref.\cite{Schoermann2013} implies Ge-concentrations between 9\,x\,10$^{19}$ and 6.2\,x\,10$^{20}$\,cm$^{-3}$ in the NDs. However, as an identical doping efficiency of NWs and NDs for small values of \textit{d}$_\text{ND}$ cannot be guaranteed we refer to the BEP$_\text{Ge}$ to describe the amount of incorporated Ge in the following sections.
Structural characterization was carried out by high-resolution transmission electron microscopy (HRTEM) and aberration corrected high angle annular dark field (HAADF) scanning transmission electron microscopy (STEM) in a FEI Tecnai F20 field emission gun microscope operated at 200\,kV and a Titan 60--300 Low Base microscope operated at 300\,kV, respectively. A PANalytical X'Pert PRO MRD setup was used for high resolution X-Ray diffraction (HRXRD) experiments.
Low temperature (4\,K) cw-ensemble PL measurements were performed using a HeCd laser operating at 325\,nm, a Jobin Yvon THR 1000 Spectrometer and a Hamamatsu R375 photomultiplier. The TR-PL analysis was carried out at 10\,K in a standard streak-camera setup \cite{Arbiol2012} with excitation by a frequency-tripled Ti:sapphire laser at 276\,nm (4.49\,eV), emitting 100\,fs pulses at a repetition rate of 78\,MHz. The streak camera provides a time resolution of 10\,ps with an overall time window of 1.5\,ns. The spectral resolution of the setup is about 5\,nm.

\begin{figure}[!tb]
  	\includegraphics[trim=0.0cm 0.0cm 0.0cm 0.0cm, width=8cm]{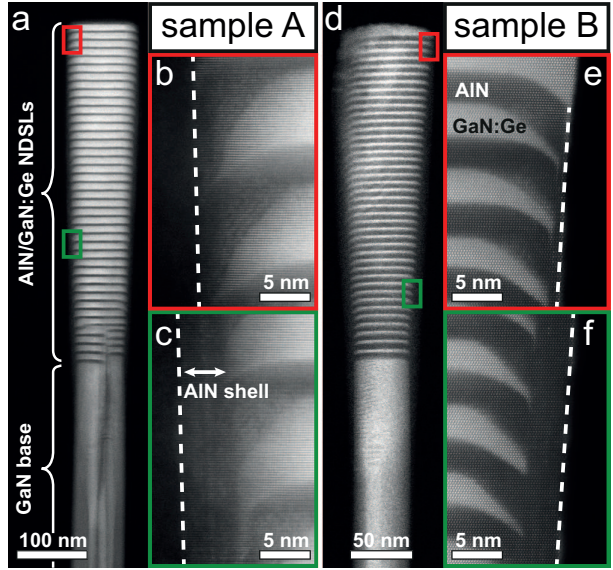}
    \caption{
Sample A: (a) Low magnification HAADF STEM image of an entire NW with GaN base and 40-fold GaN:Ge NDs (\textit{d}$_\text{ND}$\,=\,7.8\,nm, measured in the NW center) separated by AlN barriers. Bright (dark) contrast represents GaN (AlN) regions. GaN base and AlN/GaN:Ge NDSLs are indicated by braces. (b) Atomic resolution HAADF STEM image of the topmost NDs (red squared in (a)) showing a thin ($<$\,1\,nm) AlN surrounding shell. (c) Atomic resolution HAADF STEM image corresponding to the green squared area in (a) showing a well pronounced (5\,nm) AlN shell. Sample B: (d) 40-fold GaN:Ge ND system embedded in AlN barriers grown on top of a GaN NW. The lower ND growth time results in thinner NDs (\textit{d}$_\text{ND}$\,=\,2.5\,nm). (e) and (f): Atomic resolution HAADF STEM details of an area at the NW top (red rectangle in (d)) and middle section (green rectangle in (d)) of the NW, respectively.
}
\label{fig:Fig1}
\end{figure}
\begin{figure}[!tb]
  	\includegraphics[trim=0.0cm 0.0cm 0.0cm 0.0cm, width=8.5cm]{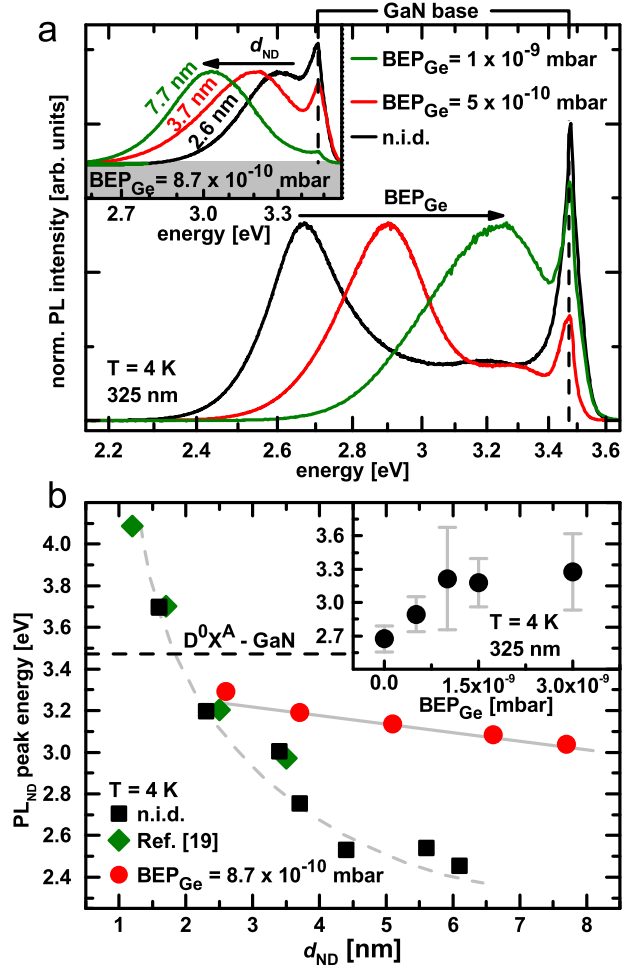}
    \caption{
(a) PL spectra of an n.i.d. and two Ge-doped samples with a nominal \textit{d}$_\text{ND}$ of 4\,nm. The NDSL PL emission blue shifts with increasing BEP$_\text{Ge}$. Inset in (a): Low temperature PL spectra of 40-fold AlN/GaN:Ge NDSLs (BEP$_\text{Ge}$\,=\,8.7\,x\,10$^{-10}$\,mbar) showing a red shift of the ND emission with increasing \textit{d}$_\text{ND}$. The signal obtained from unstrained GaN is marked by a dashed line \cite{Freitas2002}. Spectra are normalized to the NDSL emission. (b) NDSL PL peak energy as a function of \textit{d}$_\text{ND}$ for samples with different doping concentrations.  Grey lines serve as a guide to the eye, the horizontal black dashed line marks the position of the D$^\text{0}$X$^\text{A}$ in n.i.d. GaN \cite{Freitas2002}. Inset in (b): Variation of NDSL emission energy with the BEP$_\text{Ge}$. Grey vertical bars represent the FWHM of the respective emissions.    
}
\label{fig:Fig2}
\end{figure}
\begin{figure}[!tb]
  	\includegraphics[trim=0.0cm 0.0cm 0.0cm 0.0cm, width=8cm]{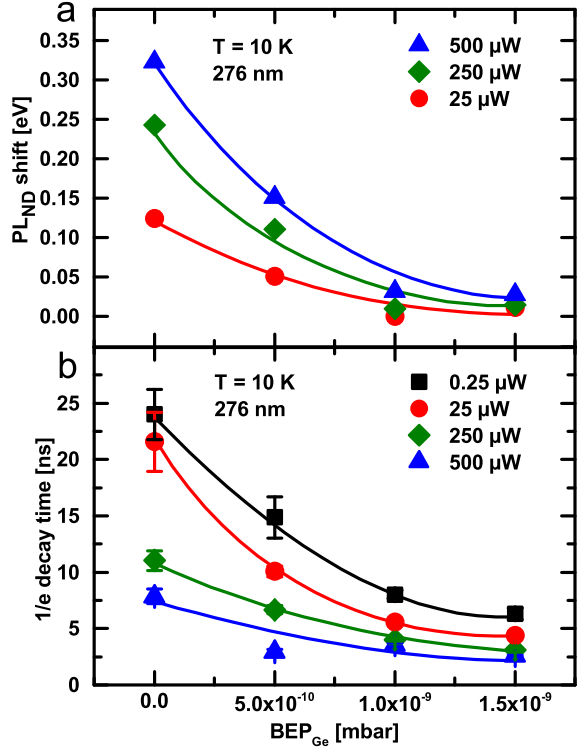}
    \caption{
(a) Shift of the NDSL PL energy relative to the emission line of 0.25\,$\upmu$W excitation for different excitation intensities as a function of the BEP$_\text{Ge}$. (b) Decay time of the NDSL PL as a function of the BEP$_\text{Ge}$ during growth for different excitation powers. Lines serve as guides to the eye.
}
\label{fig:Fig3}
\end{figure}

In \textbf{Fig.~\ref{fig:Fig1}} results of the STEM analyses of two NWHs with Ge-doped (BEP$_\text{Ge}$\,=\,8.7\,x\,10$^{-10}$ mbar) GaN NDs with different \textit{d}$_\text{ND}$ are displayed, revealing a mean thickness of the AlN barrier of (4.2$\pm$0.6)\,nm in both samples and mean values of \textit{d}$_\text{ND}$\,=\,(7.8$\pm$1.1)\,nm ((2.5$\pm$0.4)\,nm) for sample A (B), measured in the NWH center.
Images of the entire NWHs are shown in \textbf{Fig.~\ref{fig:Fig1}a} and \textbf{d}. Coalescence of adjacent NWHs and growth continuation as one NWH with an enlarged diameter is observed in numerous cases. Atomic resolution images of the NW top and middle sections are depicted in \textbf{Figs.~\ref{fig:Fig1}b,c,e,f}. The AlN barriers as well as the GaN:Ge NDs exhibit a truncated pyramidal shape introducing semipolar \{1$\overline{\text{1}}$02\} facets. No interdiffusion of Ga and Al was found in electron energy loss spectroscopy analyses, confirming the presence of sharp AlN/GaN(:Ge) interfaces (not shown). Lateral growth during barrier deposition leads to the formation of a lateral AlN shell and an increasing ND diameter in growth direction \cite{Tchernycheva2008,Furtmayr2011}. The shell thickness is less than 1\,nm at the top and up to 10\,nm at the GaN NW base (The NW outer facets are indicated by dashed lines in \textbf{Figs.~\ref{fig:Fig1}b,c,e,f}). Similar to the results reported in Ref.\cite{Furtmayr2011} the formation of misfit dislocations was observed when the ND height exceeded 3\,nm, facilitating partial strain relaxation. HRXRD analysis of the NWH ensembles revealed a high degree of out-of-plane orientation as well as a good lateral homogeneity of the NDSLs. Here, the $\omega$-2$\theta$-scan of the GaN (0002)-reflex of samples A and B exhibits clearly resolved satellite peaks up to the third order. By simulation of the diffraction patterns values for \textit{d}$_\text{ND}$ of 7.7\,nm\cite{suppl} (sample A) and 2.6\,nm (sample B) were obtained, in excellent agreement with the atomic resolution HAADF STEM results.
The results of the low-temperature cw-PL analysis are shown in \textbf{Fig.~\ref{fig:Fig2}a}. All spectra exhibit an emission line from the donor bound A exciton (D$^\text{0}$X$^\text{A}$) emission originating from the GaN base that is slightly blue shifted by up to 4\,meV with respect to the emission of unstrained GaN (3.472\,eV \cite{Freitas2002,Furtmayr2008a}) due to compressive strain induced by the AlN shell \cite{Rigutti2011,Furtmayr2011,Pierret2013}.
The energy of the emission band from the NDSL is shifted below the GaN band gap as a consequence of the QCSE due to polarization-induced internal electric fields \cite{Renard2009,Furtmayr2011}. As displayed in the inset of \textbf{Fig.~\ref{fig:Fig2}a} the AlN/GaN:Ge NDSL luminescence also red shifts with increasing \textit{d}$_\text{ND}$ for a constant BEP$_\text{Ge}$ of 8.7\,x\,10$^{-10}$\,mbar.\\
The main panel of \textbf{Fig.~\ref{fig:Fig2}a} displays low temperature PL spectra of samples with nominally \textit{d}$_\text{ND}$\,=\,4\,nm grown at different BEP$_\text{Ge}$ and a NWH sample with not intentionally doped (n.i.d.) GaN NDs for comparison. An increase of the BEP$_\text{Ge}$ results in a blue shift of the NDSL emission by up to 0.6\,eV for BEP$_\text{Ge}$\,=\,1\,x\,10$^{-9}$\,mbar (cf. inset of \textbf{Fig.~\ref{fig:Fig2}b}), strongly indicating a reduction of the QCSE due to screening of the internal electric fields by free carriers from ionized Ge donors \cite{Schoermann2013}. The spectral shape of the NDSL emission band remains almost unchanged besides an increase in its full width at half maximum (FWHM) from 250\,meV (n.i.d.) to 400\,meV (Ge-doped).\\ 
The influence of the Ge-doping on the dependence of the NDSL emission energy on \textit{d}$_\text{ND}$ is summarized in \textbf{Fig.~\ref{fig:Fig2}b}. The n.i.d. samples (black squares) show a pronounced reduction of the NDSL-related emission energy from 3.7\,eV to 2.5\,eV for an increase of \textit{d}$_\text{ND}$ from 1.6\,nm to 6.1\,nm\cite{yag}  which is in good agreement with the results obtained for nine-fold AlN/GaN ND structures from Ref.\cite{Furtmayr2011} (green diamonds). The red circles represent data obtained from AlN/GaN:Ge NDs grown with a BEP$_\text{Ge}$ of 8.7\,x\,10$^{-10}$\,mbar. In that case the red shift of the emission energy due to the QCSE is notably attenuated, again indicating a strong reduction of the internal electric fields.\\
Germanium has been shown to be an efficient donor in GaN and the BEP$_\text{Ge}$ used for the growth of GaN NDs here correspond to Ge-concentrations above 10$^{20}$\,cm$^{-3}$ in GaN NWs \cite{Hageman2004,Fritze2012,Schoermann2013}. For doping concentrations of 6\,x\,10$^{18}$\,cm$^{-3}$ screening of internal electric fields by free carriers was not observed in Si-doped GaN QWs with a thickness of \textit{d}$_\text{QW}$\,=\,3.4\,nm embedded in undoped Al$_\text{0.07}$G$_\text{0.93}$N barriers\cite{Haratizadeh2004}. Therefore, it can be concluded that noticeable screening of the internal electric fields requires a free carrier concentration above 10$^{19}$\,cm$^{-3}$. 
In order to estimate the onset of free carrier screening effects we performed 1D simulations of the carrier confinement in GaN NDs using the Schr\"{o}dinger-Poisson-solver nextnano$^3$\,\cite{nn3}. The results indicate a significant modification of the triangular ND potential, i.e. the formation of an edge potential in the ND, and a resulting blue shift of the emission above a free carrier density of $n$~$\approx$~1~x~10$^{19}$\,cm$^{-3}$ in agreement with the experimental data\cite{suppl}. These findings also agree to the theoretical work on QWs by Riyopoulos et al. \cite{Riyopoulos2009} stating that the impact of doping on the internal electric fields mainly depends on the ratio of \textit{d}$_\text{QW}$ and the Debye length $\lambda_\text{D}$. For increasing carrier density the formation of an edge potential in the QW was reported while the band profile is similar to the field-free case in the middle of the QW. At very high doping concentrations of $n$~$\approx$~10$^{20}$\,cm$^{-3}$ the transition energies approach the value of a field-free QW above the GaN band gap energy as the confined state exits the edge potential. 
Interestingly, the blue shift of the NDSL emission observed here saturates at a value of 180 meV below the GaN band gap for BEP$_\text{Ge}$~$\approx$~1\,x\,10$^{-9}$\,mbar (cf. inset in \textbf{Fig.~\ref{fig:Fig2}b}). Further increase of the Ge-concentration does not result in a further blue shift of the emission energy.
This trend was also confirmed by PL analysis for different excitation powers. In \textbf{Fig.~\ref{fig:Fig3}a} the shift of the NDSL PL emission relative to its value obtained for 0.25\,$\upmu$W excitation is displayed for different excitation intensities as a function of the BEP$_\text{Ge}$.
Whereas the n.i.d. NWHs show a strong blue shift of the NDSL emission energy with increasing excitation power due to screening of the internal electric fields by photo-excited carriers, no significant increase of the emission energy was observed for samples grown with a BEP$_\text{Ge}$ of 1\,x\,10$^{-9}$\,mbar or higher.
We assign this saturation behavior and the discrepancy between the numerical simulations and the experimental values to many body effects\cite{Helm2000} that have been neglected in Ref.\cite{Riyopoulos2009} and in our simulations.\\
In addition, the modifications of the ND confinement potential due to electrostatic screening by free carriers should result in an increasing oscillator strength and hence a decreasing PL lifetime. In order to analyze the influence of the Ge-concentration on the PL decay time TR-PL experiments at different excitation powers have been performed for samples grown with different BEP$_\text{Ge}$, the results are depicted in \textbf{Fig.~\ref{fig:Fig3}b}\cite{trpl}. A strong decrease of the PL decay time with increasing BEP$_\text{Ge}$ was found. Decay times of 24\,ns for the n.i.d. sample and of 6.3\,ns for the sample with a BEP$_\text{Ge}$ of 1.5\,x\,10$^{-9}$\,mbar were determined at a low excitation power, which is in reasonable agreement with Ref.\cite{Furtmayr2011}, reporting a PL decay time of 10\,ns for n.i.d. nine-fold AlN/GaN NDSLs (\textit{d}$_\text{ND}$\,=\,3.5\,nm) and similar excitation conditions\cite{limit}.\\
Due to the additional screening by photo-generated carriers this effect becomes less pronounced with increasing excitation power, resulting in decay times of 7.9\,ns (2.6\,ns) for the n.i.d. sample (the sample with BEP$_\text{Ge}$\,=\,1.5\,x\,10$^{-9}$\,mbar) at an excitation power of 500\,$\upmu$W. This further confirms the assignment of the decreasing PL lifetime to electrostatic screening of the internal electric fields. At the same time the evolution of the decay times also excludes a degradation of the crystal quality and the formation of non-radiative recombination centers due to the high concentration of incorporated Germanium.

In conclusion we have shown that polarization-induced internal electric fields in GaN NDs embedded in AlN/GaN NWHs can be efficiently screened by Ge-doping of the NDs. The cw-PL and TR-PL analysis unambiguously reveal that for high Ge-concentrations above 10$^{20}$\,cm$^{-3}$ the oscillator strength in the NDs is significantly increased compared to n.i.d. structures. For the highest doped samples we found a saturation behavior of the PL emission energy below the GaN band gap, indicating that due to the high density of free carriers many-body effects impact the optical transition energies.
In contrast to the case of Silicon, doping with high concentrations of Germaniun does not result in an enhanced accumulation of strain or the formation of structural defects. Consequently, high concentration of free carriers can be obtained in optically active areas. This opens new possibilities for the design of complex nanostructures for optoelectronic applications such as intersubband devices.

\begin{acknowledgments}
Authors from the Justus-Liebig-Universit\"{a}t acknowledge financial support within the LOEWE program of excellence of the Federal State of Hessen (project initiative STORE-E). The research leading to these results has received funding from the European Union Seventh Framework Programme under Grant Agreement 312483 - ESTEEM2 (Integrated Infrastructure Initiative–I3). The Marburg group gratefully acknowledges funding through the Research Training Group 1782 - Functionalization of Semiconductors by the German Science Foundation (DFG).
\end{acknowledgments}

\newpage
\setcounter{figure}{0}
\renewcommand{\thefigure}{S\arabic{figure}}
\section*{Supplemental Material}
\begin{figure}[!h]
  	\includegraphics[trim=0.0cm 0.0cm 0.0cm 0.0cm, width=8cm]{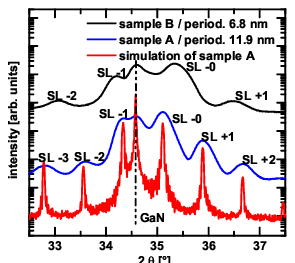}
    \caption{
HRXRD $\omega$-2$\theta$-scans of the (0002)-reflection of AlN/GaN:Ge NDSLs of samples~A (blue) and B (black). The simulation of sample~A is also shown (red) revealing a SL-period of 11.9\,nm.
}
\label{fig:FigS1}
\end{figure}

\textbf{High resolution X-Ray diffraction}
The mean value of \textit{d}$_\text{ND}$ could also be determined by high resolution X-Ray diffraction (HRXRD) analysis. \textbf{Fig.~\ref{fig:FigS1}} shows the $\omega$-2$\theta$-scan of the GaN(0002)-reflex of samples~A (blue) and sample~B (black). Both diffraction patterns exhibit clearly resolved satellite peaks up to the third order indicating a good layer periodicity. In order to assign the satellite peaks, a simulation using a dynamical diffraction theory is performed and also shown in \textbf{Fig.~\ref{fig:FigS1}} for sample~A.  Values of \textit{d}$_\text{ND}$ obtained by simulation of the diffractograms were 7.7\,nm (sample~A) and 2.6\,nm (sample~B) which is in excellent agreement with the values obtained by STEM analysis and confirms a high degree of out-of-plane orientation as well as a good lateral homogeneity of the respective thicknesses. 

\begin{figure}[!h]
  	\includegraphics[trim=0.0cm 0.0cm 0.0cm 0.0cm, width=8cm]{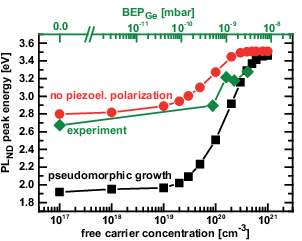}
    \caption{
Simulated transition energies of 4\,nm GaN NDs embedded in 4\,nm AlN barriers as a function of the free carrier concentration (bottom x-axis) in the GaN NDs. Black squares represent simulations assuming pseudomorphic growth, whereas full relaxation at the AlN/GaN interfaces was assumed for the data represented by red circles. Experimental transition energies in dependence on the BEP$_\text{Ge}$ (top x-axis) are represented by green diamonds.
}
\label{fig:FigS2}
\end{figure}

\textbf{Numerical simulations}
Numerical simulations of the carrier confinement were performed to estimate the influence of the free carrier concentration \textit{n} on the internal electric fields in AlN/GaN NDSLs. We have used the software package nextnano$^3$ to obtain self-consistent solutions of the 1D Schr\"{o}dinger-Poisson equation within an effective mass model\cite{nn3} using material parameters from Ref.~\cite{Vurgaftman2003}. The simulated structure consists a 10-fold NDSL with \textit{d}$_\text{ND}$\,=\,4\,nm and 4\,nm thick AlN barriers. \textbf{Fig.~\ref{fig:FigS2}} shows the dependence of the calculated transition energies on the free carrier concentration (bottom x-axis) as extracted from the confined one-particle electron and hole ground-states.
Black squares represent simulations assuming ideal pseudomorphic growth, i.e. including contributions from both spontaneous and piezoelectric polarization. Results represented by red circles neglect piezoelectric polarization (piezoelectric constant set to zero), i.e. assume full strain relaxation in the NDs to account for the formation of misfit dislocations that has been observed for ND with a thickness exceeding 3\,nm. In both cases the simulated emission energies exhibit a sigmoidal shape and a significant blue shift of the emission energy is found above \textit{n}\,=\,1~x~10$^{19}$\,cm$^{-3}$. At very high carrier concentrations of \textit{n}\,$\geq$\,10$^{20}$\,cm$^{-3}$ the transition energies converge against the value of a field-free ND in agreement with results on 2D QWs of Ref. \cite{Riyopoulos2009}. Experimental transition energies (green diamonds) are plotted as a function of BEP$_\text{Ge}$ (upper x-axis) and fall between these two cases due to partial strain relaxation at the AlN/GaN interfaces \cite{Furtmayr2011}.  As the number of activated donors is not known, the correlation between BEP$_\text{Ge}$ and $n$ is adjusted to obtain good agreement between simulation and experiment. Linear interpolation of the corresponding Ge-concentration according to Ref.\cite{Schoermann2013} yields a ratio of ionized donors of 0.88, 0.8, 0.74 and 0.66 with increasing BEP$_\text{Ge}$. A background doping concentration of 10$^{17}$\,cm$^{-3}$ has been assumed for the not intentionally doped (n.i.d.) sample. At high $n$ the simulated emission energies overestimate the experimental values, which is attributed to many body effects that have been neglected in the applied numerical model.


\begin{thebibliography}{44}
\bibitem{trpl}%
 A convolution of two exponential decays with the same time-scale but opposing time-axes and an offset of 6.4\,ns was used for fitting. This takes into account the so called back-sweep for decay-times longer than half of the repetition rate of the excitation pulses.
\bibitem{limit}%
 It has to be noted that 24\,ns is the upper detection limit of the used setup. Hence, this value should be interpreted as a lower limit and the actual decay times might be even higher.
\end{thebibliography}
\end{document}